%% file: sample-base.tex
\documentclass[10pt,conference]{IEEEtran}
\IEEEoverridecommandlockouts
\usepackage{cite}
\usepackage{amsmath,amssymb,amsfonts}
\usepackage{algorithmic}
\usepackage{graphicx}
\usepackage{textcomp}
\usepackage{xcolor}
\usepackage{booktabs} 
\usepackage{listings}
\usepackage{multirow}
\usepackage{soul}
\input{preamble} 

\definecolor{source}{gray}{0.95}
\usepackage[T1]{fontenc} 

\lstdefinelanguage{Java}{
  tabsize=4
}[keywords,comments,strings]

\definecolor{source}{gray}{0.95}
\definecolor{highlight}{gray}{0.9}

\lstnewenvironment{codesnippet}{%
	\lstset{%
		frame=single,
		framerule=0pt,
		mathescape=false
	}
}{}

\newcommand{\GH}{GitHub\xspace}
\newcommand{\SO}{Stack Overflow\xspace}

\newcommand{\rqone}{Do top crypto responders on \SO adopt cryptography in practice?\xspace}

\begin{document}
\title{Crypto Experts Advise What They Adopt}

%
%

\author{AUTHOR BLINDED}

\author{\IEEEauthorblockN{Mohammadreza Hazhirpasand}
\IEEEauthorblockN{Oscar Nierstrasz}
\IEEEauthorblockA{University of Bern\\
Bern, Switzerland}
\and
\IEEEauthorblockN{Mohammad Ghafari}
\IEEEauthorblockA{
University of Auckland\\
Auckland, New Zealand\\
m.ghafari@auckland.ac.nz}
}
\maketitle              
\begin{abstract}
Previous studies have shown that developers regularly seek advice on online forums to resolve their cryptography issues. We investigated whether users who are active in cryptography discussions also use cryptography in practice. We collected the top 1\% of responders who have participated in crypto discussions on Stack Overflow, and we manually analyzed their crypto contributions to open source projects on GitHub. We could identify 319 GitHub profiles that belonged to such crypto responders and found that 189 of them used cryptography in their projects. Further investigation revealed that the majority of analyzed users (i.e., 85\%) use the same programming languages for crypto activity on Stack Overflow and crypto contributions on GitHub. Moreover, 90\% of the analyzed users employed the same concept of cryptography in their projects as they advised about on Stack Overflow.
\end{abstract}

\section{Introduction}
\label{sec:intro}

Previous studies have shown that developers have difficulty in securely using cryptography \cite{hazhirpasand2021hurdles}, yielding many crypto misuses in software projects \cite{hazhirpasand2020java}.
Researchers have developed new tools and APIs to ease the adoption of cryptography \cite{kafader2021}, yet online Q\&A forums are among the main information sources used to resolve developer issues\cite{hazhirpasand2021survey}.

Closer inspection of experts on Q\&A forums can lead to new research directions.
For instance, profiling developer expertise contributes to heightening the members' awareness about the reliability of responses \cite{vasilescu2013stackoverflow} \cite{singer2013mutual}.
In particular, platforms such as \SO contain insecure code snippets and inexperienced developers blindly use such snippets \cite{van2020impact}.
Due to the lack of secure code examples in cryptography, we hypothesize that mapping the activity of top crypto developers cross-platform can provide an interesting path to find and evaluate their practices from the security perspective, and present such results for developers who are looking for reliable, secure crypto examples.
In this study, we conduct a preliminary step by mapping the activity of top crypto developers on \SO and \GH.
To our knowledge, no study to date has investigated the mapping of developers in cryptography across software communities.
Particularly, we address the following research question:

\emph{\textbf{RQ}: Do top crypto responders on \SO adopt cryptography in their \GH projects?}

We aim to look into the \GH profile of top 1\% of crypto responders to shed some light onto their crypto activities in practice.
We extracted the top 1\% of crypto responders (\ie 804) who participated in discussions linked to 64 cryptography tags on \SO.
We scraped their public profiles on \SO and found 319 GitHub profile links, 189 of which belonged to users who contributed to crypto files on \GH.
To assess how developers adopt cryptography in practice, we studied the \emph{programming languages} and \emph{crypto concepts} of such users across the two platforms.
We considered \emph{(1) hashing, (2) symmetric/asymmetric, (3) sign/verification} as the areas for crypto concepts.
Each of the aforementioned areas contains various algorithms and concepts. 
We realized that 85\% of analyzed users use common programming languages for crypto purposes on both platforms, 
\eg developer A resolves Java-related crypto questions on \SO, and employs Java for cryptography on \GH.
Furthermore, 90\% of the analyzed users had at least one common crypto concept on both platforms, \eg
developer A uses symmetric encryption on \GH, and helps others in the same area on \SO.
The present findings show that the practical experience of top crypto responders is noticeably in line with their theoretical experience.
Future investigations are necessary to evaluate the reliability of coding practices from the security point of view.

The remainder of this paper is structured as follows. In \autoref{sec:method}, we present the methodology of this work, then we explain the results and discuss them in \autoref{sec:rdiscussion}.
We explain the related work in \autoref{sec:related}, and explain the threats to validity of this study in \autoref{sec:threat}.
Finally, we conclude the paper in \autoref{sec:conclusion}.

\section{Methodology}
\label{sec:method}
\begin{figure*}[ht]
\centering
\centerline{\includegraphics[width=0.95\linewidth]{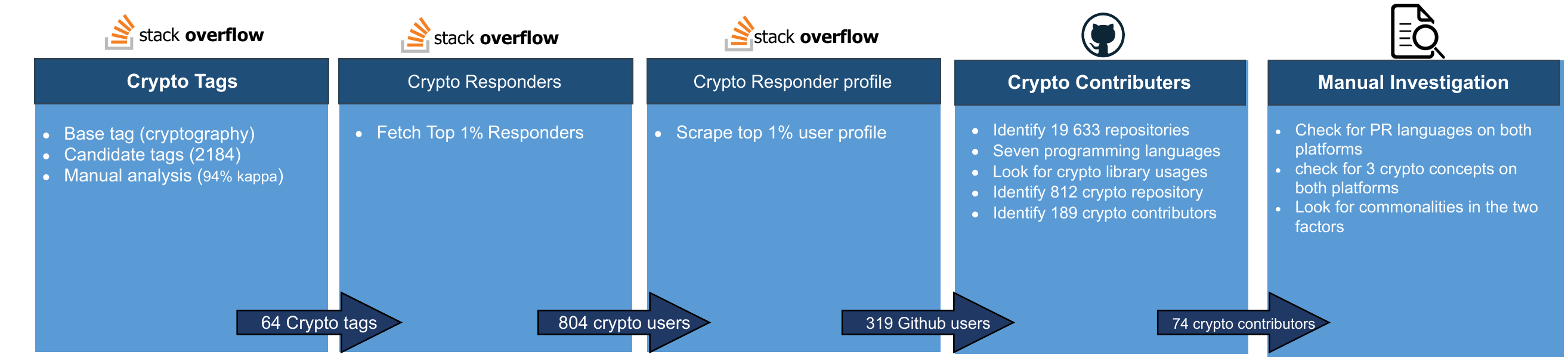}}
\caption{The pipeline for collecting and analyzing top crypto responders}\label{fig:libs}
\end{figure*}

In this section, we describe how we choose crypto tags on \SO, and our approach to fetch the top 1\% of crypto responders, extract their \GH profiles, and identify their crypto contribution (See \autoref{fig:libs}). 

\subsubsection{Crypto Tags}
To find top crypto responders on \SO, we had to identify crypto-related tags. 
We started with the ``cryptography'' tag, \ie the \emph{base tag}, to find other tags that were used together with the base tag. 
To access the data, we used the Data Explorer platform (Stack Exchange).\footnote{\url{https://data.stackexchange.com/stackoverflow/query/new}}
We found 11,130 posts that contained the base tag. 
Together with the base tag, there were 2,184 other tags, \ie \emph{candidate tags}. 
However, not all the candidate tags were related to cryptography.
The list of candidate tags is available online.\footnote{\url{http://crypto-explorer.com/mapping_data/}}

To discern crypto-related tags, two authors of this paper separately examined all the tags and marked the crypto-related ones. 
We then calculated Cohen’s kappa, a commonly used measure of inter-rater agreement \cite{cohen1960coefficient}, between the two reviewers, and achieved 94\% Cohen’s Kappa score between the two reviewers, which indicates almost perfect agreement.
Next, we compared their list of crypto tags and discussed the inconsistencies. 
Finally, we came up with a list of 64 crypto-related tags.

\subsubsection{Crypto Responders} 
We executed a query on the Data Explorer platform to fetch the top 1\% of crypto responders for each of the identified tags from \SO.
\autoref{tab:cryptotags} presents the 64 tags and associated top 1\% of \emph{unique} crypto responders. 
We excluded the crypto responders that we had already found in other tags.
For instance, the crypto++ tag had four top crypto responders, considering that they were among other tags.
In total, we retrieved 804 top crypto responders.
The list of top crypto responders is available online.\footnote{\url{http://crypto-explorer.com/mapping_data/}}

\begin{table}[]
\scriptsize
\centering
\caption {The 64 crypto tags and associated unique top 1\% crypto responders (\ie 804)} \label{tab:cryptotags}
\begin{tabular}{rl|rl}
\hline
\textbf{Responders}          & \multicolumn{1}{l|}{\textbf{Tag}} & \textbf{Responders} & \multicolumn{1}{l}{\textbf{Tag}} \\ \hline
202	&	 encryption	&	2	&	encryption-asymmetric	\\
176	&	hash	&	2	&	cryptoapi	\\
98	&	cryptography	&	2	&	pbkdf2	\\
76	&	openssl	&	2	&	jca	\\
29	&	md5	&	2	&	jasypt	\\
20	&	keystore	&	2	&	commoncrypto	\\
16	&	xor	&	2	&	libsodium	\\
14	&	digital-sig	&	2	&	phpseclib	\\
13	&	sha1	&	1	&	ellipticurve	\\
12	&	x509certificate	&	1	&	ecdsa	\\
11	&	rsa	&	1	&	diffie-hellman	\\
10	&	mcrypt	&	1	&	rsacryptoserviceprovider	\\
8	&	sha256	&	1	&	bcrypt	\\
8	&	private-key	&	1	&	node-crypto	\\
8	&	sha	&	1	&	sjcl	\\
8	&	public-key	&	1	&	spongycastle	\\
7	&	bouncycastle	&	1	&	cryptoswift	\\
7	&	smartcard	&	1	&	hashlib	\\
6	&	public-key-encryption	&	1	&	wolfssl	\\
5	&	x509	&	0	&	crypto++	\\
5	&	salt	&	0	&	pkcs11	\\
5	&	hmac	&	0	&	jce	\\
5	&	pycrypto	&	0	&	pkcs7	\\
4	&	cryptojs	&	0	&	cng	\\
4	&	pyopenssl	&	0	&	cryptographic-hash-function	\\
3	&	aes	&	0	&	aescryptoserviceprovider	\\
3	&	encryption-symmetric	&	0	&	rijndaelmanaged	\\
3	&	rijndael	&	0	&	webcrypto-api	\\
3	&	3des	&	0	&	mscapi	\\
3	&	m2crypto	&	0	&	charm-crypto	\\
3	&	botan	&	0	&	javax.crypto	\\
2	&	des	&	0	&	nacl-cryptography	\\ \hline
\end{tabular}
\end{table}

\subsubsection{Crypto Responder Profile}
\SO offers the ability to its users to share their social media addresses (\eg Twitter, GitHub, and personal websites) on their profile. 
Nevertheless, the aforementioned information is not accessible on Stack Exchange Data Explorer.
Hence, to find the selected users' \GH profiles, we automatically scraped profiles of the 804 Stack Overflow top crypto responders.
Using the BeautifulSoup library in Python, we parsed each user profile automatically. 
For 804 \SO users, we could identify 319 \GH profiles. 

\subsubsection{Crypto Contributors}
We used the \GH repository API and collected a total of 19\,633 public repositories associated with the 319 \GH users.
We selected the top seven programming languages used in the repositories, \ie Python, Ruby, C, C++, Rust, Java, and C\#. 

To understand which crypto libraries are popular in the selected languages, we consulted with two crypto experts.
Among the suggested names, there are some candidates that come with the languages, such as \emph{Java.security} in Java, or the libraries that are widely accepted and well-known, such as \emph{Bouncy Castle} for Java and C\#.
Afterward, to ensure the rest of the suggested libraries are largely accepted in developer community, we checked how popular (\ie star and fork) the suggested open-source crypto libraries are on GitHub, \eg \emph{libsodium} for the C language had 9.2k stars and 1.4k forks.
The crypto libraries had on average 1844 stars and 346 forks, and the median number were 1105 and 245, respectively.

Using the compiled list of crypto libraries in \autoref{tab:cryptolib}, we employed the \GH Code Search API and a custom regex script to identify in which files crypto namespaces, \eg ``System.Security'', were used.
At the time of writing this paper, the \GH Code Search API could not perform the exact keyword search for the crypto namespaces. 
Therefore, we relied on a supplementary regex script to ensure the identified code snippets contain the namespaces.
We retrieved a total of 2\,404 crypto files in 812 repositories.

In the last step, we used \emph{git blame} to identify the contributors who had committed to the 2\,404 crypto files.
To do so, we cloned the 812 crypto repositories and extracted authors and committers of crypto files by git blame.
We then fetched the developers' email addresses, usernames, and full names by \GH user API in order to check whether they are among the contributors of the 812 crypto repositories.
Of the 319 top crypto responders on \SO, we found that 189 developers had crypto contributions on \GH.
They had on average 14 and 3 median crypto file contributions.

\begin{table*}[ht]
\small
\centering
\caption {The selected crypto libraries in the seven programming languages} \label{tab:cryptolib} 
\begin{tabular}{lllllll}
\hline
\textbf{Python} & \textbf{Ruby} & \textbf{Java} & \textbf{C}        & \textbf{C++} & \textbf{C\#}                 & \textbf{Rust} \\
\hline
passlib         & bcrypt-ruby   & Java.security & libgcrypt         & Botan        & Bouncy Castle                & octavo        \\
pynacl          & Ruby Themis   & Javax.crypto  & NaCl          & Cryptlib     & libsodium-net                & rustls        \\
hashlib         & digest        & Bouncy Castle & crypto-algorithms & Cryptopp     & system.security.cryptography & rust-crypto   \\
pythemis          & RbNaCl        &               & Themis            & HElib        & PCLCrypto                    & sodiumoxide   \\
PyElliptic      &               &               & wolfSSL           &           &                              & crypto        \\
bcrypt          &               &               & libsodium         &              &                              & Ring     \\
          &               &               & S2N-tls         &              &                              &          
    
\end{tabular}
\end{table*}

\subsubsection{Manual Investigation}
To address the research question, we performed a manual analysis to observe to what extent users employ cryptography in practice.
To this end, we checked two aspects of their contribution, \emph{(1) the programming language used for crypto purposes on both platforms, (2) crypto concepts used on both platforms}.

Identifying detailed crypto concepts in various crypto libraries as well as crypto discussions can be an arduous task.
Therefore, we deduced the concepts used in this study from recent work on the categorization of developers' crypto challenges on \SO \cite{hazhirpasand2021hurdles}.
The researchers' findings revealed that developers mostly encounter challenges concerning hashing, symmetric/asymmetric, and digital signature.
Accordingly, we assumed that developers commonly use three high-level crypto concepts, which are \emph{(1) hashing, (2) symmetric/asymmetric, and (3) signing/verification}.

In our manual analysis, we attempted to find commonalities in
the programming languages (\ie the seven languages) and crypto concepts that are used by a developer on both platforms.

To compute the sample size for studying 189 users on \GH, we defined a confidence level of 95\% and 9\% as the margin of error, which yields 74 for our sample size.
We then randomly selected 74 users from the population.
Writing queries on the Stack Exchange Data Explorer platform, we automatically retrieved all the posts (\ie titles, question and answer body) wherein the 74 developers were involved on \SO.

Two authors of this paper manually reviewed all the posts to extract the programming languages used in the discussions, \ie question and answer body.
Afterward, they also checked the title and question body to understand to which concept or concepts a particular discussion can be assigned. 
They checked the crypto codes of the 74 users on \GH, and extracted the crypto concept(s), and recorded the programming language of the crypto files. 
To understanding the crypto concepts, they looked for the APIs used in the crypto files. 
For instance, if the MessageDigest API was used in a Java crypto file, they assumed that the developer encountered the hashing topic in practice.
In cases where they had doubts about the APIs, they referred to the API documentation of the library.
They had several sessions in order to compare the results of their investigations and build a unified list.
Ultimately, they checked for commonalities of the languages and the crypto concepts that the users used across the two platforms.

\section{Results and Discussion}
\label{sec:rdiscussion}
In this section, we present and discuss our findings for the following research question: \emph{\rqone}
We explore the usage of crypto responders' programming languages and crypto concepts on \SO and \GH.


\subsubsection{\SO} We extracted 804 top crypto responders in which 319 users shared their \GH profile on \SO.
We fetched the crypto discussions of the 74 users (the sample size), extracted their provided answers, and stored the names of the programming languages involved in the discussions.
In total, 55\% of discussions were about Java. 
A user could have participated in various discussions wherein different programming languages were involved. 
We therefore considered all those languages as being the areas of the user's crypto knowledge. 
The median value of programming languages used on \SO is 3 and 2.7 is the average value.

More than four-fifths of the developers (\ie 65) participated in discussions where the three crypto-concepts were discussed.
Similar to programming languages, a user can provide answers for a discussion in which the knowledge of a concept or mixed concepts are required.
For instance, we considered \emph{(1) hashing (2) sign/verification} for the discussion (ID:33305800) on \SO since a user was confused about the differences between hashing with SHA256 and signing with SHA256withRSA.

\subsubsection{\GH} 
Of 319 users with \GH profiles, 189 had made crypto contributions to public repositories on \GH.
To conduct our manual analysis, we randomly selected 74 users from the 189 crypto developers.
We extracted the names of programming languages where crypto APIs were used.
The median value of programming languages used on \GH is 1 and the average value is 1.4.
In all 74 cases, the number of programming languages and crypto concepts on \SO was higher than or equal to the same groups of data on \GH.
For instance, developer A participated in discussions where three languages (\ie C++, C\#, Java) were involved as well as the three crypto concepts while the same developer only used Java crypto APIs for hashing purposes on \GH.

\begin{figure}[ht]
\centering
\centerline{\includegraphics[width=1\linewidth,trim=4 4 4 4,clip]{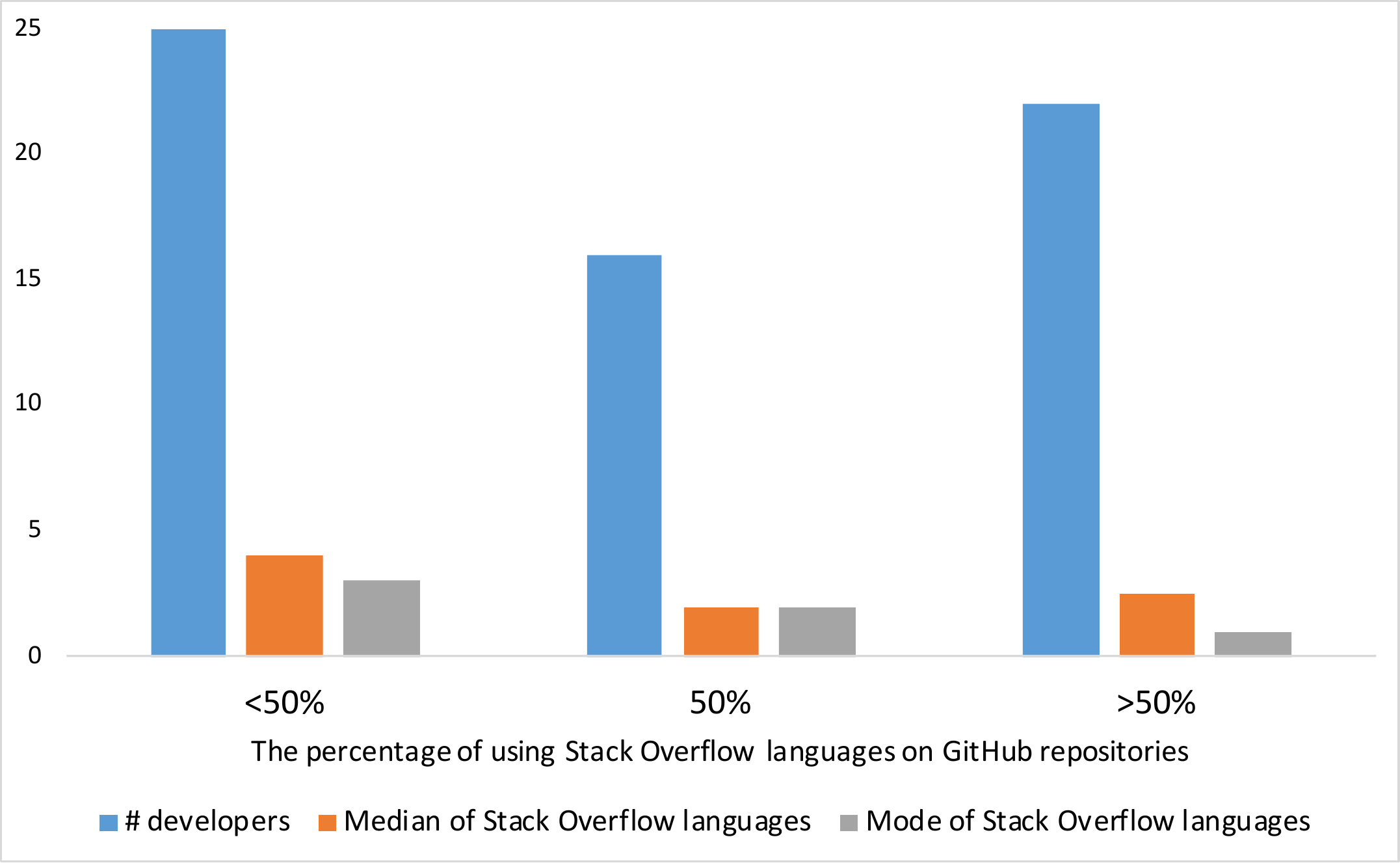}}
\caption{The number of developers based on their percentage of \SO programming languages usage in \GH repositories}\label{fig:sotogh}
\end{figure}

\subsubsection{Mapping result}
Interestingly, we realized that 63 (\ie 85\%) of such users had used at least one language that matches their crypto activity on \SO.
Such agreement implies that the users are confident in those languages.
We split the 63 developers into three groups: those who used fewer than 50\% of the languages in their \GH open-source projects (\ie 25), those who used half of the languages (\ie 16), and those using more than 50\% of the languages (\ie 22) (See \autoref{fig:sotogh}).
In particular, more than half of the developers (\ie 38) had crypto contributions for either half or more than half of the languages that they prefer to provide crypto help for on \SO.
The developers who used fewer than 50\% of their \SO languages in open-source projects constitute 39\% (\ie 25) of the whole.

\begin{figure}[ht]
\centering
\centerline{\includegraphics[width=1\linewidth,trim=4 4 4 4,clip]{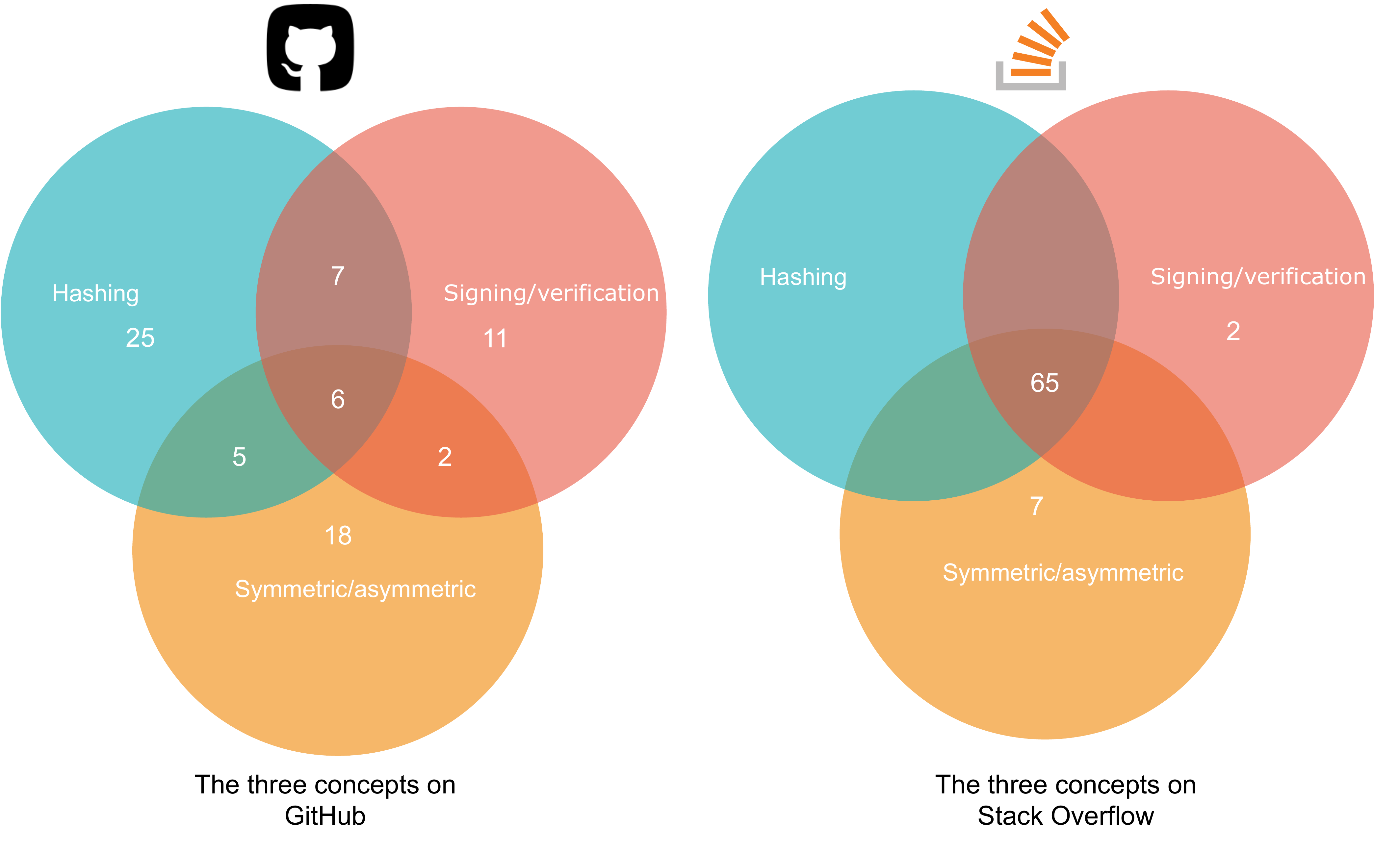}}
\caption{The number of developers in each crypto concept on \SO and \GH}\label{fig:concepts}
\end{figure}

With regard to crypto concepts, there are 6 developers who used APIs on \GH which are related to the three crypto concepts (See \autoref{fig:concepts}). 
There are seven developers who used \emph{signing/verification} and \emph{hashing}, five developers who employed \emph{hashing} and \emph{symmetric/asymmetric}, and only two developers used \emph{signing/verfication} and \emph{symmetric/asymmetric}. 
The rest of developers only used one of the concepts in the identified projects. 
They might have a broader contribution to cryptography in open-source projects, however, it may be due to the limitation of our obtained knowledge concerning their practices on \GH.
On the other side, the manual investigation revealed that, on \SO, 65 developers participated in all three concepts, seven developers only in symmetric/asymmetric, and only two in signing/verification.
Checking the labels of 74 developers, we uncovered that almost all of the developers (\ie 67 or 90\% ) worked with at least one common crypto concept on both platforms.
Of the 67 users, 30\% of them had more than one concept shared on both platforms.
The findings imply that developers are confident in programming languages and the crypto concepts as they had relevant experience in practice.
Likewise, user satisfaction, such as high upvotes for the responses on \SO, confirm that the users' guidance is practical and effective in the domain of cryptography.

\section{Threats to validity}
\label{sec:threat}
We identified 804 developers who were among the top 1\% of responders to 64 crypto tags on \SO. 
However, we were only able to find 319 of these developers on \GH, and did not perform any exhaustive search on Google to find more users.
A developer may have multiple accounts on \GH for various purposes but we only consider one account per user.
Some users may have private repositories and make more significant contributions to crypto-related projects, nevertheless, such contributions cannot be assessed.
We looked into the repositories written in seven programming languages, and did not analyze the remaining repositories.
Even though we included popular and default crypto libraries in each programming language, adding more crypto libraries in each programming language can allow a more realistic conclusion to be drawn.
This is important, considering that the diversity of crypto libraries in each language is debatable.
We used the git blame command to fetch a crypto file's contributors. 
Consequently, there is a likelihood that the developers who contributed to crypto files had committed to other parts of the file but not to the cryptography parts.

\section{Related work}
\label{sec:related}
The significance of correctly employing cryptography and obtaining professional help from online sources has been discussed by numerous authors in the literature.
Sifat \etal studied three popular online sources, \ie crypto Stack Exchange,  Security Stack Exchange, and Quora, to find out the common challenges concerning implementing security in data transmission \cite{jahan2017exploratory}.
Yang \etal carried out a large-scale analysis of security-related questions on \SO and reported a classification of five topics \cite{yang2016security}. 
A recent study conducted by Meng \etal has recognized the challenges of writing secure Java code on \SO \cite{meng2018secure}.
Their results provide compelling evidence to the fact that the security implications of coding options in Java, \eg CSRF tokens, are partially grasped by many developers.
Lastly, a study confirmed that developers are uncritically using the insecure code snippets found on \SO \cite{fischer2017stack}.
The aforementioned findings jeopardize the security of software\cite{fischer2019stack}.
We observed that relying on poorly validated responses on online forums was inextricably linked to software systems' security implications.
In this research, we studied the crypto experts who frequently help others on \SO to observe if they adopt cryptography in practice.

A series of recent studies have focused on profiling developer expertise either on single or multiple platforms \cite{yan2018profiling} \cite{bouguessa2008identifying}.
A common concern in profiling developer expertise cross-platforms is to track developer identity, as developer activity can be dispersed from one platform to another \cite{kouters2012s}.
For instance, Zhang \etal used the developer email and the hashing approach to identify the same developer with the same email address on another platform \cite{zhang2017devrec}.
Yung \etal looked into the challenge of expert finding with the Topic Expertise Model (TEM) \cite{liucqarank}. 
Their approach jointly modeled topics and expertise by combining textual content model and link structure analysis.
Tian \etal proposed a novel methodology to extract experts that utilizes various user attributes and related platform-specific information, for instance, high-quality Stack Overflow answers in specific programming technologies and high-quality projects measured using source code metrics \cite{tian2019geek}.

Sajedi \etal checked the features that overlap between GitHub and Stack Overflow \cite{badashian2014involvement}. 
They  defined three high-order metrics related to both networks (\ie development, management and popularity)
Their findings reveal moderate and strong correlations between the derived metrics within each platform.
Vasilescu \etal analyzed the differences of 46,967 active users both on \SO  and \GH to understand the \SO's involvement of the \GH's developers \cite{vasilescu2013stackoverflow}.
They discovered that users who provide more answers on \SO tend to have a high number of commits.
Their results imply that users with a high number of commits on \GH have a greater tendency to take the role of a ``teacher'' instead of asking more questions on \SO.
Vadlamani \etal focused on perceiving what constitutes the notion of an expert developer and what key elements affect developer contribution\cite{vadlamani2020studying}. 
They conducted a survey with active software developers both on \SO and \GH.
Their results show that developers consider personal drivers to be more critical than professional factors for \GH contribution, and the majority of experts participate in both private and public repositories. Furthermore, developers do not seem to be willing to participate on \SO as the questions are either uninteresting or easy, and they find the reward system demotivating.

\section{conclusion and future plans}
\label{sec:conclusion}
We conducted a study of the top 1\% of crypto responders on \SO to shed some light onto the adoption of cryptography on \GH by the top crypto responders on \SO. 
In particular, to the best of our knowledge, no previous study has profiled crypto developers across online communities.
We found 189 users who used cryptography in open-source projects on \GH and studied 74 of this population.
The results indicate that the majority of analyzed users (\ie 85\%) use the same programming languages for participating in crypto discussions on \SO and crypto contributions on \GH.
Closer inspection of three areas in cryptography (\ie hashing, symmetric/asymmetric, or signing/verification) revealed that 90\% of the analyzed users had practical experience with at least one of the crypto concepts that they had discussed on \SO.
Collectively, the results demonstrate that top crypto users are consistent with their crypto activity on both platforms, and this provides a basis for further research to investigate the quality of their practical experience.

\section{Acknowledgments}
We gratefully acknowledge the financial support of the Swiss National Science Foundation for the project
``Agile Software Assistance'' (SNSF project No.\ 200020-181973, Feb.\ 1, 2019 - April 30, 2022).


\bibliographystyle{IEEEtran}
\bibliography{sample-base}

\end{document}

%% file: preamble.tex
\usepackage{xspace}
\usepackage{graphicx}
\usepackage{ifthen}
\usepackage[normalem]{ulem} 
\usepackage{xcolor,colortbl}
\usepackage{hyperref}

\newboolean{showedits}
\setboolean{showedits}{true} 
\ifthenelse{\boolean{showedits}}
{
	\newcommand{\del}[1]{\textcolor{red}{\sout{#1}}} 
	\newcommand{\nbe}[3]{
		{\colorbox{#3}{\bfseries\sffamily\scriptsize\textcolor{white}{#1}}}
		{\textcolor{#3}{\sf\small$\blacktriangleright$\textit{#2}$\blacktriangleleft$}}}
}{
	\newcommand{\del}[1]{} 
	
	\newcommand{\nbe}[3]{}
}

\newboolean{showcomments}
\setboolean{showcomments}{true} 
\newcommand{\id}[1]{$-$Id: scgPaper.tex 32478 2010-04-29 09:11:32Z oscar $-$}

\ifthenelse{\boolean{showcomments}}
 {
 	\newcommand{\nbc}[3]{
 		{\colorbox{#3}{\bfseries\sffamily\scriptsize\textcolor{white}{#1}}}
		{\textcolor{#3}{\sf\small$\blacktriangleright$\textit{#2}$\blacktriangleleft$}}}
	
 }{
 	\newcommand{\nbc}[3]{}
 	
 }

\usepackage[most]{tcolorbox}
\ifthenelse{\boolean{showedits}}
{
  \newtcolorbox{inserted}{%
       title=Inserted text:,
       colframe=blue,colback=blue!5!white,
       breakable,
       leftrule=0mm, 
       bottomrule=0mm,
       rightrule=0mm,
       toprule=0mm,
       arc=0mm, outer arc=0mm,
       oversize
  }
  \newtcolorbox{deleted}{%
       title=Deleted text:,
       colframe=red,colback=red!5!white,
       breakable,
       leftrule=0mm, 
       bottomrule=0mm,
       rightrule=0mm,
       toprule=0mm,
       arc=0mm, outer arc=0mm,
       oversize
  }
  \newtcolorbox{refactored}{%
       title=Rewritten text:,
       colframe=blue,colback=red!5!white,
       breakable,
       leftrule=0mm, 
       bottomrule=0mm,
       rightrule=0mm,
       toprule=0mm,
       arc=0mm, outer arc=0mm,
       oversize
  }
}{

}
\newboolean{isblinded}
\setboolean{isblinded}{true}
\ifthenelse{\boolean{isblinded}}
{\newcommand\blind[1]{BLINDED\xspace}}
{\newcommand\blind[1]{#1\xspace}}

\newcommand{\commented}[1]{}

\newcommand{\eg}{\emph{e.g.,}\xspace}
\newcommand{\ie}{\emph{i.e.,}\xspace}
\newcommand{\etal}{\emph{et al.}\xspace}
